\documentclass[review]{elsarticle}
\usepackage{graphicx}
\usepackage{amsfonts}
\usepackage{amssymb}
\usepackage{amsmath}

\journal{PHYSICA A}

\begin{document}

\begin{frontmatter}

\title{Thermodynamic model of social influence\\ on two-dimensional square lattice:\\
Case for two features}

\author{Jozef Genzor}
\author{Vladim\'{\i}r Bu\v{z}ek}
\author{Andrej Gendiar\corref{mycorrespondingauthor}}

\address{Institute of Physics, Slovak Academy of Sciences, SK-845~11, Bratislava,
Slovakia}

\cortext[mycorrespondingauthor]{Corresponding author}
\ead{andrej.gendiar@savba.sk}

\begin{abstract}
We propose a thermodynamic multi-state spin model in order to describe
equilibrial behavior of a society. Our model is inspired by the Axelrod
model used in social network studies. In the framework of the statistical
mechanics language, we analyze phase transitions of our model, in which
the spin interaction $J$ is interpreted as a mutual communication among
individuals forming a society. The thermal fluctuations introduce a noise
$T$ into the communication, which suppresses long-range correlations. Below
a certain phase transition point $T_t$, large-scale clusters of the individuals,
who share a specific dominant property, are formed. The measure of the cluster
sizes is an order parameter after spontaneous symmetry breaking.
By means of the Corner transfer matrix renormalization group algorithm,
we treat our model in the thermodynamic limit and classify the phase
transitions with respect to inherent degrees of freedom. Each individual
is chosen to possess two independent features $f=2$ and each feature can
assume one of $q$ traits (e.g. interests). Hence, each individual is described by
$q^2$ degrees of freedom. A single first order phase transition is detected in
our model if $q>2$, whereas two distinct continuous phase transitions are found
if $q=2$ only. Evaluating the free energy, order parameters, specific heat,
and the entanglement von Neumann entropy, we classify the phase transitions
$T_t(q)$ in detail. The permanent existence of the ordered phase (the
large-scale cluster formation with a non-zero order parameter) is conjectured
below a non-zero transition point $T_t(q)\approx0.5$ in the asymptotic regime
$q\to\infty$.
\end{abstract}


\end{frontmatter}


\section{Introduction}

The original Axelrod model~\cite{Axelrod}, despite its simplicity, has
been accepted as a model describing social influence with respect to
the interaction among individuals~\cite{bbv,Castellano}. The interacting
individuals of the Axelrod model are located on a regular two-dimensional
square lattice and are characterized by a set of $f$ cultural features,
whereas each feature can assume one of the $q$ cultural traits. Each
feature represents one of the cultural
dimension, e.g., language, religion, technology, style of dress, etc.
The dynamics of this model runs in two steps repeated as required. In the 
first step, an agent (representing an individual) and one of his neighbors 
are selected randomly. In the second step, the probability of interaction
is calculated being proportional to the number of shared features.  
The interaction consists of selecting a feature at random in which the two
agents differ, and setting this feature of the neighboring agent to be equal
to the corresponding feature of the agent. It might seem that this mechanism
leads to homogenization of society. However, it can lead either to a global
homogenization (the ordered phase) or to a fragmented state (disordered phase)
with coexistence of different homogeneous regions. One of the main results
conjectures a critical value $q_c$ separating a monocultural state from
the multicultural~\cite{Cast2}. Further details of the thermodynamic
Axelrod model and the out-of-equilibrium Axelrod model are discussed
in Ref.~\cite{Gandica}.

In this study we consider a classical multi-spin model of a social
system treated from the point of view of the statistical mechanics.
We focus our attention on
behavior of a model of a large society in equilibrium. The society is
represented by individuals who mutually interact via communication
channels (e.g. sharing interests) with the nearest neighbors
only. The society is subject to special rules given by a model
of the statistical mechanics we have introduced for this purpose.
A noise plays an important role in this study. The noise interferes
with the communication channels. If it increases, the communicating
individuals get less correlated on larger distances. In this way the
noise acts against the formation of larger clusters of the individuals
with a particular character, i.e., a set of shared features. In such
a cluster, the individuals share a similar social background. The size
of the clusters can be quantified by calculations of an appropriate
order parameter, correlation length, etc., which are commonly used in
the statistical physics. If a phase transition point exists in a given
statistical model, this point separates an ordered phase from the
disordered. The two phases can be determined by the order parameter
being non-zero within the ordered phase or zero in the disordered,
provided that the system is infinitely (sufficiently) large, and the
spontaneous symmetry-breaking mechanism occurred below the transition
point. The noise can be also regarded as random perturbations (cultural
drift) realized as a spontaneous change in a trait~\cite{Toral1} and can
play a significant constructive role in the out-of-equilibrium Axelrod model.
On the other hand, the effect of the noise for such non-linear dynamical
systems is found to be size-dependent~\cite{Toral2}.

We, therefore, propose a multi-state spin model on the two-dimensional
regular square lattice of the infinite size. Each vertex of the lattice contains 
a multi-state spin variable (being an individual with
a certain cultural setting). We define special nearest-neighbor
interactions among the spins representing a conditional communication
among individuals. The statistical Gibbs distribution introduces
thermal fluctuations into our model with a multi-spin Hamiltonian.
Here, the temperature can be identified as the noise we introduced above.
Imposing a constant magnetic field on given spin states makes the spins
align accordingly, which might have had a similar effect as, for instance,
the mass media or advertisement. Having calculated the effects of the
magnetic field, we observed a typical paramagnetic response in our model
only, and no phase transition was observed.

The model describes thermodynamic features of social influence studied by
the well-known Axelrod model~\cite{Axelrod}. Gandica {\it et al.}~\cite{Gandica}
have recently studied such thermodynamics features in the coupled Potts models
in one-dimensional lattices, where the phase transition occurs at zero
temperature in accord with a thermodynamic one-dimensional interacting
multi-state spin system. Our studies go beyond this thermodynamic Axelrod
model conjectures since we intend to study phase transitions on social
systems at non-zero temperature, where number of the individuals is
infinite. Therefore, the spontaneous symmetry-breaking mechanism
selects a certain preferred cultural character resulting in a large cluster
formation, which is characterized by a non-zero order parameter.

This task is certainly nontrivial since our model has not been known to
have an analytical solution. Therefore, we apply the Corner Transfer Matrix
Renormalization Group (CTMRG) algorithm~\cite{ctmrg1}, which is a powerful
numerical tool in the statistical mechanics. The CTMRG calculates
all thermodynamic functions to a high accuracy and enables to analyze
the phase transitions as well as to control the spontaneous symmetry
breaking. The phase transition temperature decreases with increasing
number of traits $q$ as discussed later. We intend to
investigate the asymptotic case in this paper, i.e., the case when the
number of the traits $q$ of each individual is infinite. Then, we estimate
the phase transition point in order to find out whether the ordered phase
is permanently present or not. In other words, the phase transition point
$T_t$ is found to remain non-zero. Throughout this paper we consider the
case of $f=2$ only.

The paper is organized as follows. In Sec.~II we define the Hamiltonian
of our model, briefly describe CTMRG, and introduce the thermodynamic
functions used in the analysis of the phase transition. The Sec.~III
contains numerical calculations explained in the statistical physics
language. In Sec.~IV we discuss and interpret our results in terms of
the communicating individuals influenced by the presence of the noise.

\section{Lattice model and CTMRG algorithm}

\subsection{Hamiltonian and density matrix}

A classical spin lattice model is considered on the regular two-dimensional
square lattice, where the nearest-neighbor multi-state spins placed on
the lattice vertices interact. Let $\sigma_{i,j} = 0, 1, \dots, n-1$ be a
generalized multi-spin with integer degrees of freedom $n$. The subscript indices
$i$ and $j$ denote the position of each lattice vertex, where
the spins are placed within the $X$ and $Y$ coordinate system on the underlying
lattice, i.e., $-\infty<i,j<\infty$. We start with the $n$-state clock
(vector) model~\cite{clock} for this purpose with the Hamiltonian
\begin{equation} \label{Clock}
{\cal H} = - J \sum\limits_{i=-\infty}^{\infty}
               \sum\limits_{j=-\infty}^{\infty}
               \sum\limits_{k=0}^{1}
                     \cos(\theta_{i,j} - \theta_{i+k,j-k+1})\, .
\end{equation}
The interaction term $J$ acts between the nearest-neighbor vector spins
$\theta_{i,j} = 2\pi\sigma_{i,j}/n$. The $k$ summation includes the
horizontal and the vertical directions on the square lattice.

Let us generalize the spin clock model so that the interaction term $J$
contains a special attribute, i.e., extra spins are added. We, therefore,
introduce additional degrees of freedom to each vertex. The Hamiltonian in
Eq.~\eqref{Clock} can be further modified into the form ${\cal H}=\sum_{ijk}
J_{ijk}\cos\,(\theta_{i,j} - \theta_{i+k,j-k+1})$. The position dependent
term $J_{ijk}$ describes the spin interactions $J$ of the $n$-state clock
model controlled by additional $q$-state Potts model
$\delta$-interactions~\cite{FYWu}. The total number of the spin degrees of
the freedom is $nq$ on each vertex $i,j$. We study the simplified case when
$q\equiv n$ starting from the case of $q=2$ up to $q=6$ which is still
computationally feasible. (In more general case when $q\neq n$,
we do not expect substantially different physical consequences as
those studied in this work.)

Hence, our multi-state spin model contains two $q$-state spins on
the same vertex, i.e., 
$\sigma_{i,j}^{(1)}=0,1,2,\dots,q-1$ and $\sigma_{i,j}^{(2)}=0,1,2,\dots,q-1$,
which are distinguished by the superscripts ($1$) and ($2$). It is instructive to
introduce a $q^2$-variable $\xi_{i,j} = q\sigma_{i,j}^{(1)} + \sigma_{i,j}^{(2)}
= 0,1,\dots,q^2-1$. The Hamiltonian of our model has its final form
\begin{equation}
\label{Final_H}
{\cal H} = \sum\limits_{i,j=-\infty}^{\infty}
            \sum\limits_{k=0}^{1}
               \left\{ J^{(1)}_{ijk}
                  \cos\left[\theta^{\left(2\right)}_{i,j}
                          - \theta^{\left(2\right)}_{i+k,j-k+1}
                      \right]
              +   J^{(2)}_{ijk}
                  \cos\left[\theta^{\left(1\right)}_{i,j}
                          - \theta^{\left(1\right)}_{i+k,j-k+1}
                      \right]
              \right\},
\end{equation}
noticing that $\theta_{i,j}^{(\alpha)} = 2\pi\sigma_{i,j}^{(\alpha)}/q$, where
\begin{equation}
J^{(\alpha)}_{ijk} =
      -J\delta\left(\sigma^{(\alpha)}_{i,j},\,
                    \sigma^{(\alpha)}_{i+k,j-k+1}
              \right)
 \equiv
  \begin{cases}
    -J, & \text{if\ \ \ } \sigma^{(\alpha)}_{i,j} = 
                    \sigma^{(\alpha)}_{i+k,j-k+1},\\
  \phantom{-}0, & \text{otherwise}.
  \end{cases}
\end{equation}
The superscript $(\alpha)$ can take only two values as mentioned above.
The Potts-like interaction $J_{ijk}^{(\alpha)}$ is represented by a diagonal
$q\times q$ matrix with the elements $-J$ on the diagonal.

Thus defined model can also describe conditionally communicating (interacting)
individuals of a society. The society is modeled by individuals ($\xi_{i,j}$)
and each individual has two distinguished features $\sigma^{(1)}$ and
$\sigma^{(2)}$. Each feature assumes $q$ different values (traits).
In particular, an individual positioned on $\{i,j\}$ vertex of the square
lattice communicates with a nearest neighbor, say $\{i+1,j\}$, by comparing
the spin values of the first feature $\sigma^{(1)}$. This comparison is
carried out by means of the $q$-state Potts interaction. If the Potts
interaction is non-zero, the individuals communicate via the $q$-state
clock interaction of the other feature with
$\alpha=2$. The cosine enables a broader communication spectrum than the
Potts term. Since we require symmetry in the {\em Potts-clock} conditional
communication, we include the other term in the Hamiltonian, which
exchanges the role of the features ($1$) and ($2$) in our model.
In particular, the Potts-like communication first compares the feature
$J_{ijk}^{(2)}$ followed by the cosine term with the feature $\alpha=1$.
(Enabling extra interactions between the two features within each individual
and/or the cross-interactions of the two adjacent individuals is to be studied
elsewhere.) The total number of all the individuals is considered to be
infinite in order to detect and analyze the phase transition when the
spontaneous symmetry breaking is present.

In the framework of the statistical mechanics, we investigate a combined
$q$-state Potts and $q$-state clock model which is abbreviated as the
$q^2$-state spin model. As an example, one can interpret the case of $q=3$
in the following: the feature $\sigma^{(1)}$ can be chosen
to represent {\em leisure-time interests} while the other feature $\sigma^{(2)}$
can involve {\em working duties}. In the former case, one could list three
properties such as reading books, listening to music, and hiking, whereas the
latter feature could consist of manual activities, intellectual activities,
and creative activities, as the example. The thermal fluctuations, induced by
the thermodynamic temperature $T$ of the Gibbs distribution, are meant to
describe a noise hindering the communication. The higher the noise, the
stronger suppression of the communication is resulted.

We classify the phase transitions of our model by numerical calculation of the
partition function ${\cal Z}$
\begin{equation}
  \label{part_fnc}
  {\cal Z}=\sum\limits_{\{\sigma\}} \exp\left(-\frac{{\cal H}}{k_B T}\right)\, ,
\end{equation}
especially, by its derivatives. The sum has to be taken through all
multi-spin configurations $\{\sigma\}$ on the infinite lattice. Here, Boltzmann
constant and temperature are denoted by $k_B$ and $T$, respectively. The
partition function is evaluated numerically by the CTMRG algorithm~\cite{ctmrg1},
which generalizes the Density Matrix Renormalization Group~\cite{White} on the
two-dimensional classical spin systems. In the CTMRG language, the whole square
lattice is divided into four identical quarters (corners of the square shape),
the so-called corner transfer matrices, and the renormalization group (RG)
transformation projects out all those spin configurations which have the lowest
probability selected by a density matrix.

A typical formulation of an observable (an averaged thermodynamic function)
$\langle\hat X\rangle$ obeys the standard expression
\begin{equation}
  \langle\hat X\rangle = {\cal Z}^{-1} \sum\limits_{\{\sigma\}}
     \hat X\exp\left(-\frac{{\cal H}\{\sigma\}}{k_B T}\right)
     \equiv {\rm Tr}_s \left( \hat X \hat \rho_s \right)\, ,
\end{equation}
where the matrix $\hat\rho_s$ is introduced being commonly called the
{\em reduced density matrix}
\begin{equation}
 \label{rdm}
  \hat \rho_s = {\cal Z}^{-1} \sum\limits_{\{\sigma_e\}}
       \exp\left(-\frac{{\cal H}\{\sigma\}}{k_B T}\right)\, .
\end{equation}
It is a classical counterpart of the one-dimensional quantum reduced density
matrix in DMRG defined for a subsystem $s$ in contact with an environment $e$.
The reduced density matrix is defined on a line of the spins $\{\sigma_s\}$
(forming the subsystem $s$) between any of the two adjacent corner transfer
matrices, whereas all the remaining spins variables form the environment $e$.
The configuration sum is taken over all spins within the environment $\{\sigma_e\}$
except those of the subsystem $\{\sigma_s\}$. Notice the normalization
${\rm Tr}_s \hat \rho_s = 1$. Its meaning is the partition function ${\cal Z}$
within the classical statistical physics and is normalized to unity.

Our model can be thought of as a system with two non-trivially coupled
sub-lattices, where either sub-lattice is composed of the $q$-state
variables with the given feature $\alpha$.

\subsection{Thermodynamic functions}

The {\em Helmholtz free energy} $F$ per spin site
\begin{equation}
F = - k_{\rm B}T\ln\left({\cal Z}\right)
\end{equation}
can be easily evaluated from the partition function by CTMRG. Taking derivatives
of the free energy determines other thermodynamic functions used in the
classification of the phase transition. Namely, the first derivative with
respect to temperature $T$ results in the
{\em internal energy}
\begin{equation}
 \label{int_eng}
U = - T^2 \dfrac{\partial \left(F/T\right)}{\partial T}\, ,
\end{equation}
which is equivalent to the nearest-neighbor correlation function
evaluated on the square lattice for the Potts-like models
\begin{equation}
 \label{nn-corl}
U = - J \langle \sigma_{i,j}\sigma_{i,j+1} \rangle
    - J \langle \sigma_{i,j}\sigma_{i+1,j} \rangle\, .
\end{equation}
The consequent derivative of the internal energy with respect to
$T$ yields the {\em specific heat}
\begin{equation}
 \label{spec_heat}
C = \dfrac{\partial U}{\partial T}\, ,
\end{equation}
which has a non-analytic (divergent) behavior at a phase transition.
Analogously, the first and the second derivatives of the free energy
with respect to an external magnetic field $h$ results in the magnetization
(the order parameter) and the susceptibility, respectively. Another important
thermodynamic function to be calculated is the {\em entanglement von Neumann
entropy} $S_v$. It follows the standard quantum-mechanical definition
\begin{equation}
\label{ee}
S_v = -\text{Tr}_s\left(\hat{\rho}_s \log_2 \hat{\rho}_s \right).
\end{equation}
This quantity reflects the correlation effects, which are maximal at the
phase transition point.

The order parameter $\langle O\rangle$ can be equivalently evaluated
via the reduced density matrix in Eq.~\eqref{rdm} being either
non-zero within an ordered spin phase or zero in the disordered.
A continuous transition usually leads to the second-order phase transition,
and the discontinuous behavior signals the first-order phase transition.
However, a detailed analysis of the free energy and other thermodynamic
functions is usually necessary to distinguish the order of the phase transition.

Let us define a {\em sub-site} order parameter for a given feature
$\alpha$
\begin{equation} \label{order1}
\langle O_{\alpha}\rangle = {\rm Tr}_s\left(\hat{O}_s^{(\alpha)}\hat{\rho}_s\right)
= {\rm Tr}_s\left[\cos\left(\frac{2\pi\sigma^{(\alpha)}_{i,j}}{q}\right)
\hat{\rho}_s\right]\, ,
\end{equation}
where the sub-site order parameter $\hat O_s^{(\alpha)}$ is measured. For
simplicity, we excluded the subscripts ${i,j}$ from the order parameter
notation. Another useful definition of the order parameter, measuring both
of the spins at the same vertex, is a {\em complete} order parameter
\begin{equation} \label{order2}
\langle O\rangle = {\rm Tr}_s\left(\hat{O}_s\hat{\rho}_s\right)
 = {\rm Tr}_s\left[\cos\left(2\pi\frac{\xi_{i,j}-\phi}{q^2}\right)
\hat{\rho}_s\right]\, .
\end{equation}
Again, we simplified the expression into $\xi = q\sigma^{(1)} + \sigma^{(2)}$.
We also extended the definition of
the complete order parameter by introducing a $q^2$-state fixed parameter $\phi$.
This parameter $\phi$ specifies the alignment of $\langle O\rangle$ towards a
reference spin level, where the multi-state spin projections are measured.
Unless stated explicitly in the text, we often consider the parameter $\phi=0$.

The CTMRG algorithm has been a well-established numerical method for almost
two decades and recognized by the physical community as an accurate and
reliable numerical method~\cite{Uli1,Uli2}. All of the thermodynamic functions
can be calculated to a high accuracy, which is governed by the
integer number of the CTMRG/DMRG states kept $m$ (the higher the number $m$,
the better accuracy is reached~\cite{ctmrg1,White}). Throughout this work we
used $100\leq m\leq 200$, which led to the RG truncation error~\cite{White}
as small as $\varepsilon \lesssim 10^{-8}$ around the phase transition,
otherwise the error reaches the machine precision. Additional increasing of the states
kept $m$ does not change our results.

\begin{figure}[!tb]
\centerline{\includegraphics[width=0.95\textwidth,clip]{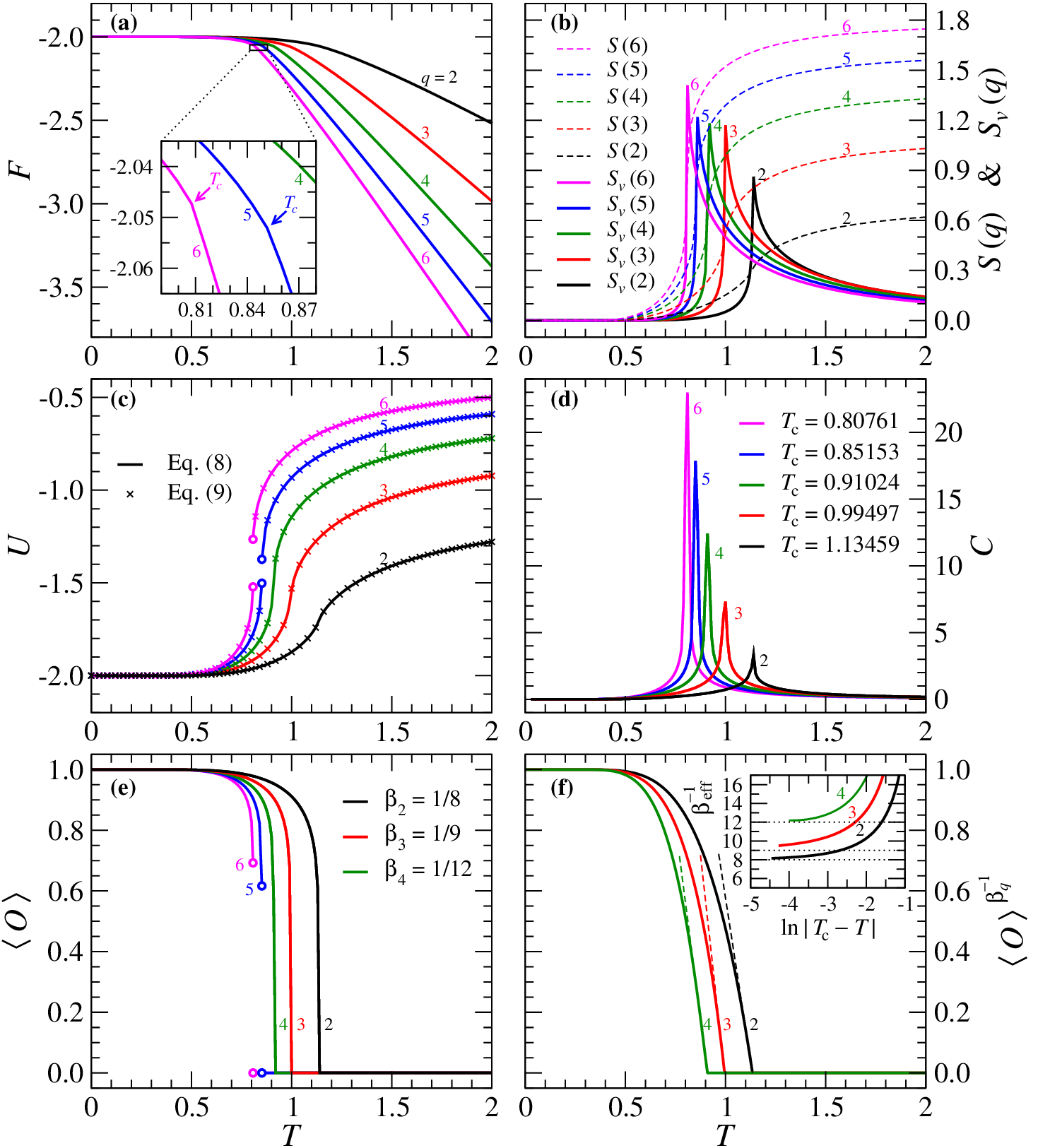}}
\caption{(Color online) The temperature dependences of the thermodynamic functions
in the $q$-state Potts model obtained by CTMRG:
(a) the free energy $F$, here the inset depicts the details of the two kinks when
the first order transition appears, (b) the thermodynamic entropy $S(q)=-\frac{
\partial F}{\partial T}$ and the entanglement entropy $S_v(q)$ from Eq.~\eqref{ee},
(c) the internal energy $U$ obtained by Eqs.~\eqref{int_eng} (full lines) and
\eqref{nn-corl} (cross symbols), (d) the specific heat $C$ from Eq.~\eqref{spec_heat},
(e) the complete order parameter $\langle O\rangle$ from Eq.~\eqref{order2} if $\phi=0$,
and (f) the inverse power $\beta_q^{-1}$ to the complete order parameter exhibiting
the linear dependences right below the critical point (dashed lines); the inset shows
the detailed analysis of the effective exponent $\beta_{\rm eff}$ with respect to
the limit $T \to T_c$ as in Eq.~\eqref{beff}.}
\label{fig1}
\end{figure}

In order to test the efficacy of the CTMRG, we study the standard $q$-state
Potts model with a constant magnetic field $h$. The field $h$ requires an additional
term $-h\delta\left(\sigma_{i,j}^{(k)},0\right)$ in the Hamiltonian~\eqref{Final_H}.
Since the $q$-state Potts models have analytic expressions for the phase transition
temperature $T_c^{-1}(q)=\ln(1+\sqrt{q})$, the numerical calculations carried out on
this model serves as a benchmark for our model studied later. Figure~\ref{fig1}
depicts a couple of selected thermodynamic functions of the $q$-state Potts model
for $q=2,3,...,6$. We calculated the free energy per site $F$ in the panel (a). The
model exhibits the second order phase transition for $2\leq q\leq 4$. The first
order transition is present if $q\geq5$, and the free energy has a non-analytic kink
as depicted in the inset of the panel (a). The thermodynamic entropy $S(q)=-\frac
{\partial F}{\partial T}$ and the von Neumann entanglement entropy $S_v(q)$, cf.
Eq.~\eqref{ee}, are shown in the panel (b). The phase transition temperatures
$T_c(q)$ correspond to the $S_v(q)$ maxima. The equivalence of the internal energies
$U$ obtained by Eqs.~\eqref{int_eng} (full lines) and \eqref{nn-corl} (the symbols
{\small$\times$}) are plotted in the panel (c). Here, the discontinuities appearing
in $U$ are proportional to the non-zero latent heat, which unambiguously confirm the
presence of the first order phase transition for $q=5$ and $6$. The divergent peaks
of the specific heat $C$ in the panel (d) coincide with the phase transition
temperatures $T_c(q)$. The height of the peaks will increase if a finer temperature
sampling of $U$ is used; each height is limited by the numerical derivatives of the
internal energy within the given sampling. The panel (e) shows the complete order
parameter $\langle O\rangle$ remains continuous in the second order transition, but
exhibits a discontinuous jump which is typical for the first order phase transition.
We also evaluated the critical exponent $\beta_q$, which are related to the order
parameter $\langle O\rangle \propto |T_c(q)-T|^{\beta_q}$. These critical exponents
are in agreement with the analytical solutions~\cite{FYWu}. The linearity of the
order parameter $\langle O\rangle^{\beta_q^{-1}}$ when temperature approaches the
critical point from the ordered phase are depicted in the panel (f), where the
dashed lines are the tangents at $T_c(q)$. The inset shows, the convergence of the
effective magnetic exponent
\begin{equation}
 \label{beff}
 \beta_{\rm eff}(T,q) = \frac{\partial \ln \langle O(T,q)\rangle}
                             {\partial \ln \left[T_c(q) - T\right]}\, ,
\end{equation}
where $\beta_q=\lim\limits_{T\to T_c(q)}\beta_{\rm eff}(T,q)$. The horizontal dotted
lines serve as guides for the eyes, which correspond to the critical exponents
$\beta_q$ reached in the asymptotic limit $\ln\left[T_c(q)-T\right]\to-\infty$.

\begin{figure}[tb]
\centerline{\includegraphics[width=0.75\textwidth,clip]{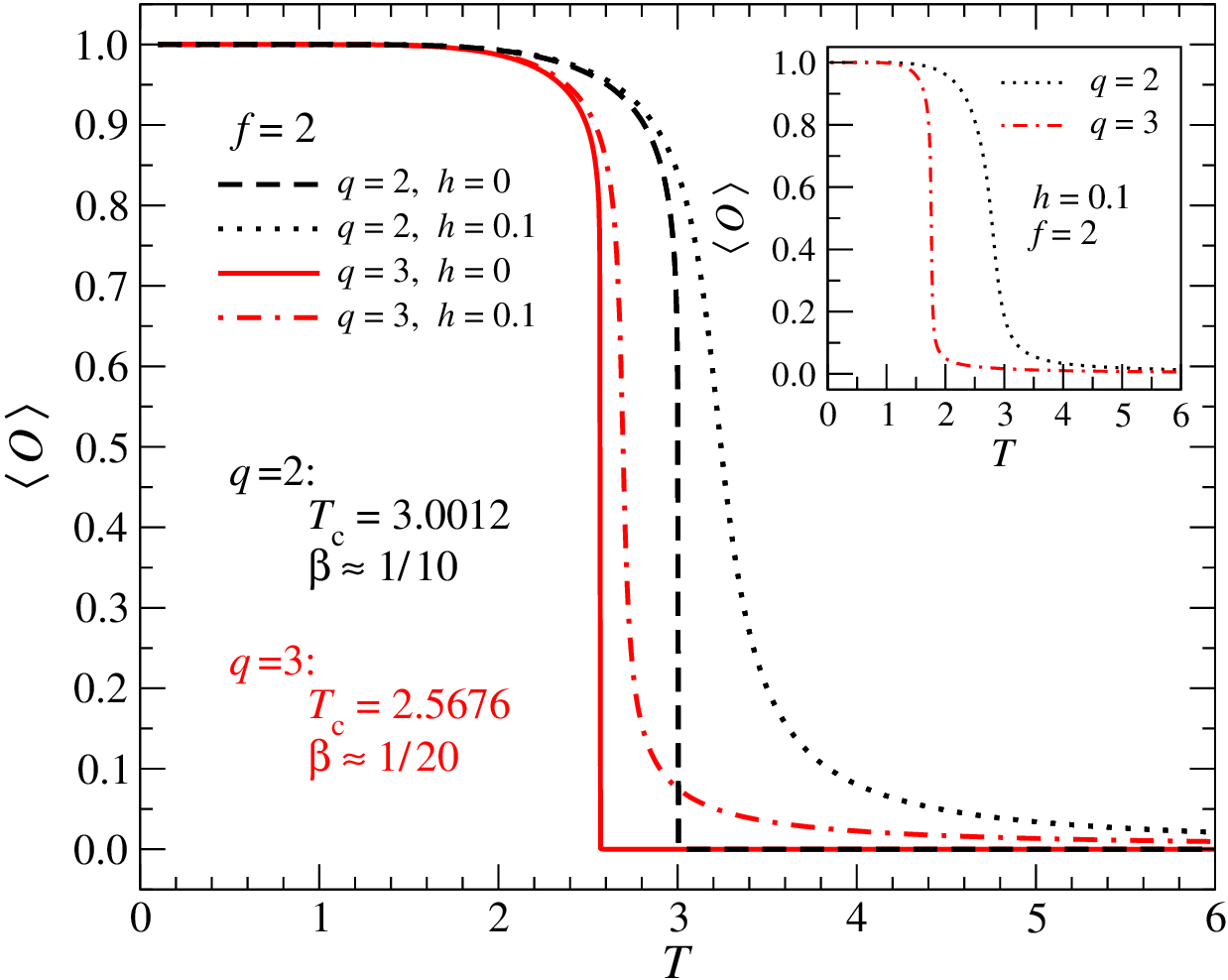}}
\caption{(Color online) The temperature dependence of the complete order
parameter $\langle O\rangle$ on the two-dimensional square lattice of the
thermodynamic version of the Axelrod model studied in Ref.~\cite{Gandica}
in the case of $f=2$. A typical response of the model on the magnetic field
$h$ is shown for $q=2$ and $q=3$. (The inset depicts supplemental information
on our model Hamiltonian in Eq.\eqref{Final_H} if the magnetic field $h=0.1$
is imposed.)}
\label{fig2}
\end{figure}

If the magnetic field $h$ is set to be non-zero, the thermodynamic functions are
always analytic within all temperature range, and no phase transition point is
detected. Figure~2 shows this case for $h=0$ and $h=0.1$ if we applied CTMRG to
the model Hamiltonian studied in Ref.~\cite{Gandica} on the two-dimensional
square lattice. It is evident that for zero field the dashed ($q=2$) and the
full ($q=3$) lines exhibit the continuous phase transitions with the critical
temperatures and exponents $T_c=3.0012$, $\beta\approx\frac{1}{10}$ and
$T_c=2.5676$, $\beta\approx\frac{1}{20}$, respectively. Applying the magnetic
field $h=0.1$, the phase transition is not present, and the model responds in
the standard paramagnetic way for $q=2$ (dotted line) and $q=3$ (the dashed-dotted
line). Since we are interested in the phase transition analysis of our model,
we exclude detailed analysis with non-zero magnetic field in our model.

\section{Numerical results}

\begin{figure}[tb]
\centerline{\includegraphics[width=0.75\textwidth,clip]{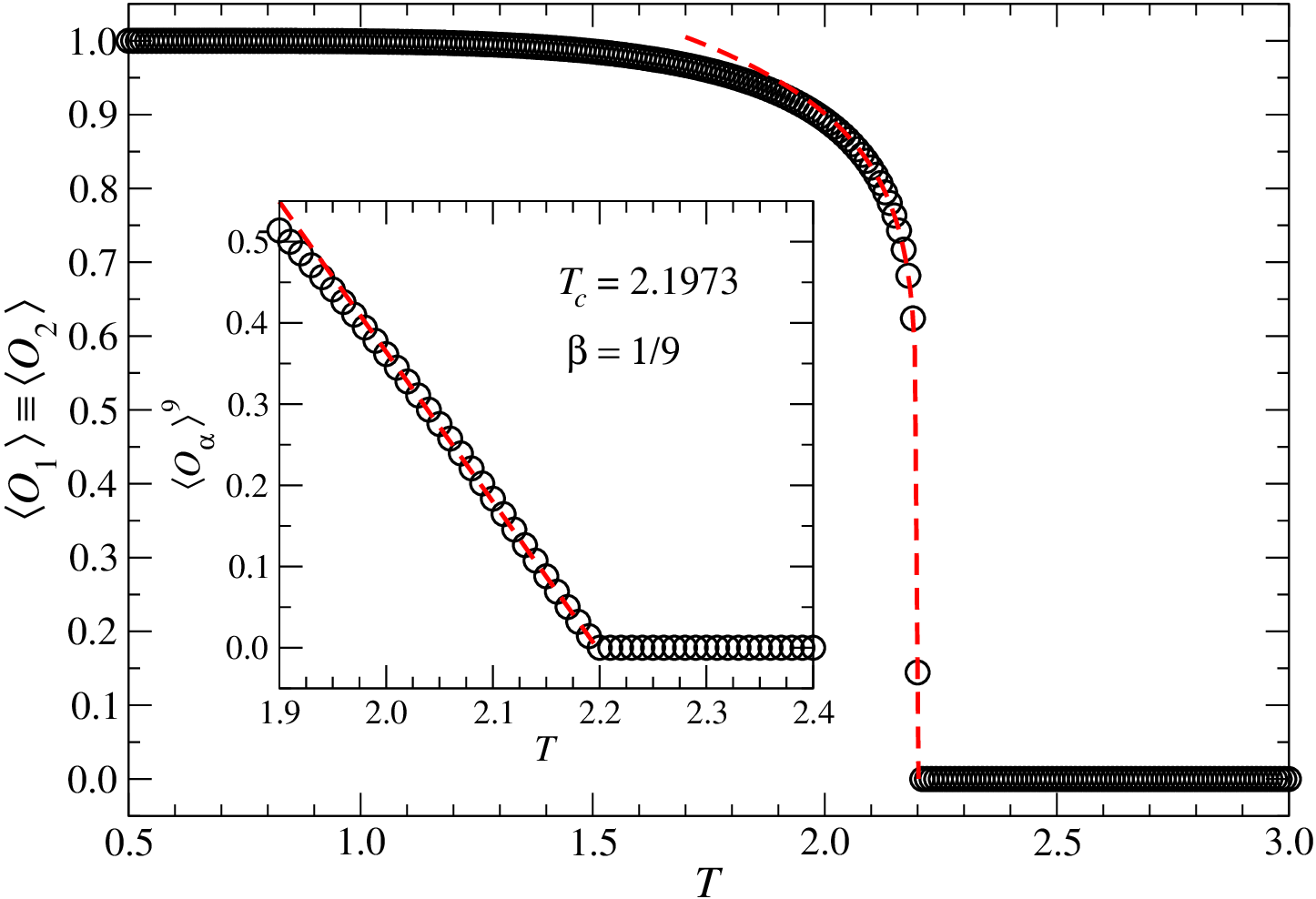}}
\caption{(Color online) The temperature dependence of the sub-site order
parameter $\langle O_{\alpha}\rangle$ (black circles) for $f=2$ and $q=2$
remains unchanged for both $\alpha = 1, 2$. The red dashed line corresponds
to the scaling relation with the critical exponent $\beta\approx0.1113$,
which was analyzed in the same way as shown in the panel (f) of Fig.~\ref{fig1}.
The inset depicts the ninth power of $\langle O_{\alpha}\rangle$ with the
expected linearity below $T_c$.}
\label{fig3}
\end{figure}

The phase transitions in the classical spin systems are induced by the thermal
fluctuations by varying the temperature $T$ in Eq.~\eqref{part_fnc}. We use
dimensionless units, in which $J = k_{\rm B} = 1$.  This corresponds to
the ferromagnetic spin ordering. We begin with the simplest non-trivial
case of $q=2$. Figure~\ref{fig3} shows the sub-site order parameter $\langle
O_{\alpha}\rangle$ with respect to temperature $T$ which is identical for both
$\alpha=1$ and $\alpha=2$. The second order phase transition is resulted at the
critical temperature $T_c = 2.1973$. The associated universality scaling
$\langle O_{\alpha}\rangle\propto\left(T-T_c\right)^{\beta}$ results in the
common critical exponent $\beta\approx0.1113$. The inset shows nearly linear
behavior of $\langle O_{\alpha}\rangle^{1/\beta}$ when approaching the
critical temperature $T$ from the ferromagnetic phase. The critical exponent
of our model at $q=2$ is very close to the $3$-state Potts model universality
class~\cite{FYWu}, where $\beta=\frac{1}{9}$. This model analogy is non-trivial
and requires further clarification. Notice that the exponent $\beta\approx
0.1113$ differs from the well-known Ising ($2$-state clock) universality,
where $\beta=\frac{1}{8}$. It belongs neither to the $4$-state Potts nor
$4$-state clock model universality classes.

\begin{figure}[tb]
\centerline{\includegraphics[width=0.75\textwidth,clip]{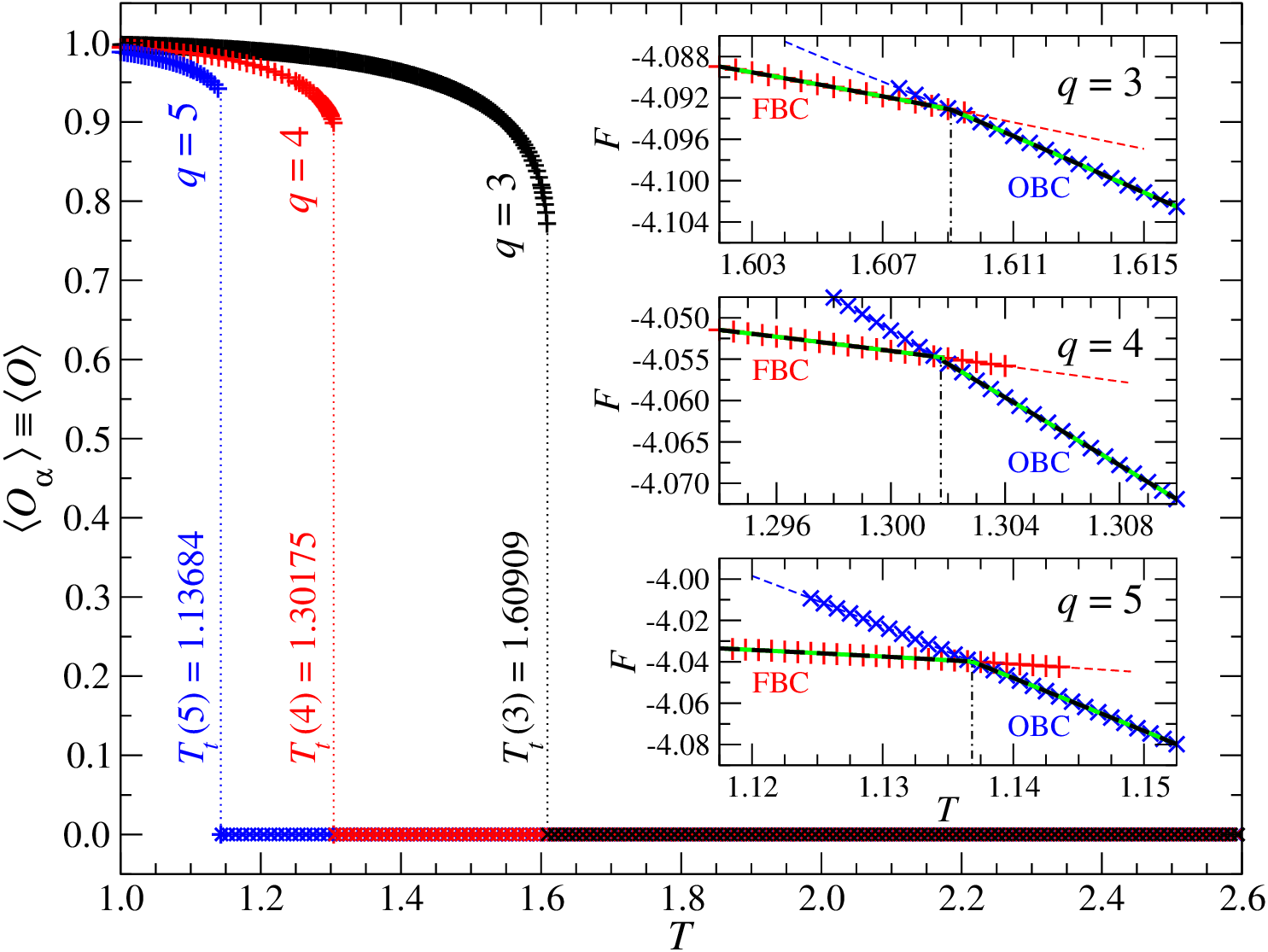}}
\caption{(Color online) The order parameter $\langle O_{\alpha}\rangle$
is discontinuous for $q\geq3$ ($f=2$) and reflects the first order phase
transition in all three cases when $3\leq q\leq5$. The free energy $F$ is
depicted in the respective three insets. We compare $F$ for the fixed BCs
(red symbols) and for the open BCs (blue symbols) around the transition
temperature.}
\label{fig4}
\end{figure}

The sub-site order parameter $\langle O_{\alpha}\rangle$ for $q=3$, $4$, and
$5$ is depicted in Fig.~\ref{fig4}. It gradually decreases with increasing
temperature, but at certain temperature it discontinuously jumps to zero.
Such behavior usually suggests the first order phase transition. To confirm
this statement, the free energy $F$ per spin is plotted with respect to $T$
for two different boundary conditions (BCs). The fixed (open) BCs are imposed
at the very beginning of the iterative CTMRG scheme in order to enhance
(suppress) spontaneous symmetry breaking resulting in the ordered (disordered)
phase in a small vicinity of the phase transition point. In particular, if
the fixed BCs are applied, the spontaneous symmetry-breaking mechanism
selects one of $q^2$ free energy minima as specified by the fixed BCs.
On the contrary, the open BCs prevent the spontaneous symmetry breaking from
falling into a minimum and makes the system be in a metastable state below
the phase transition. Since the first-order phase transition is known to exhibit
the coexistence of two phases in a small temperature interval around the phase
transition, such an analysis with the two different BCs is inevitable to locate
the phase transition accurately. The insets for the three cases, $q=3,4,5$,
show the normalized Helmholtz free energy around the transition temperature.
The red and blue symbols of the free energy correspond to the
fixed and the open BCs, respectively. The temperature interval, in which two distinguishable
converged free energy are measured according to BCs set, is the region, where
the ordered and disordered phases can coexist. The true phase transition
point $T_t(q)$ is located at the free energy crossover, and the equilibrium free
energy is shown by the thick dashed line corresponding to the lower free energy.
In this case, the free energy is a non-analytical at $T_t(q>2)$ and exhibits
a kink typical for the first-order phase transition (further details on the
first order analysis are can be found in Ref.~\cite{3dPotts}). Taking the
derivatives of $F$ with respect to $T$, a discontinuity of the thermodynamic
functions in Eqs.~\eqref{int_eng} and \eqref{spec_heat} is resulted. (We remark
here that the free energy is not sensitive to the different BCs if a critical
second-order phase transition is present, i.e., if $q=2$.)

The phase transition temperatures for $q>2$ are calculated within a high
accuracy resulting $T_t(3)=1.60909$, $T_t(4)=1.30175$, $T_t(5)=1.12684$,
and $T_t(6)=1.03234$ (not plotted) at the crossing point of the free energy.
It is obvious that $T_t(q)$ gradually decreases with increasing $q$, and later
we study the asymptotic case when $q\to\infty$. It is also worth to mention
that the first-order phase transition is not critical in sense of the non-diverging
correlation length at the phase transition temperature (not shown) in contrast
to the second order phase transition, when the correlation length diverges.
For this reason, we reserve the term {\em critical} temperature, $T_c(q)$, for
the second-order phase transition only, which is resulted in our model only
if $q=2$. Otherwise, we use the notation {\em transition} temperature $T_t(q)$.

\begin{figure}[tb]
\centerline{\includegraphics[width=0.75\textwidth,clip]{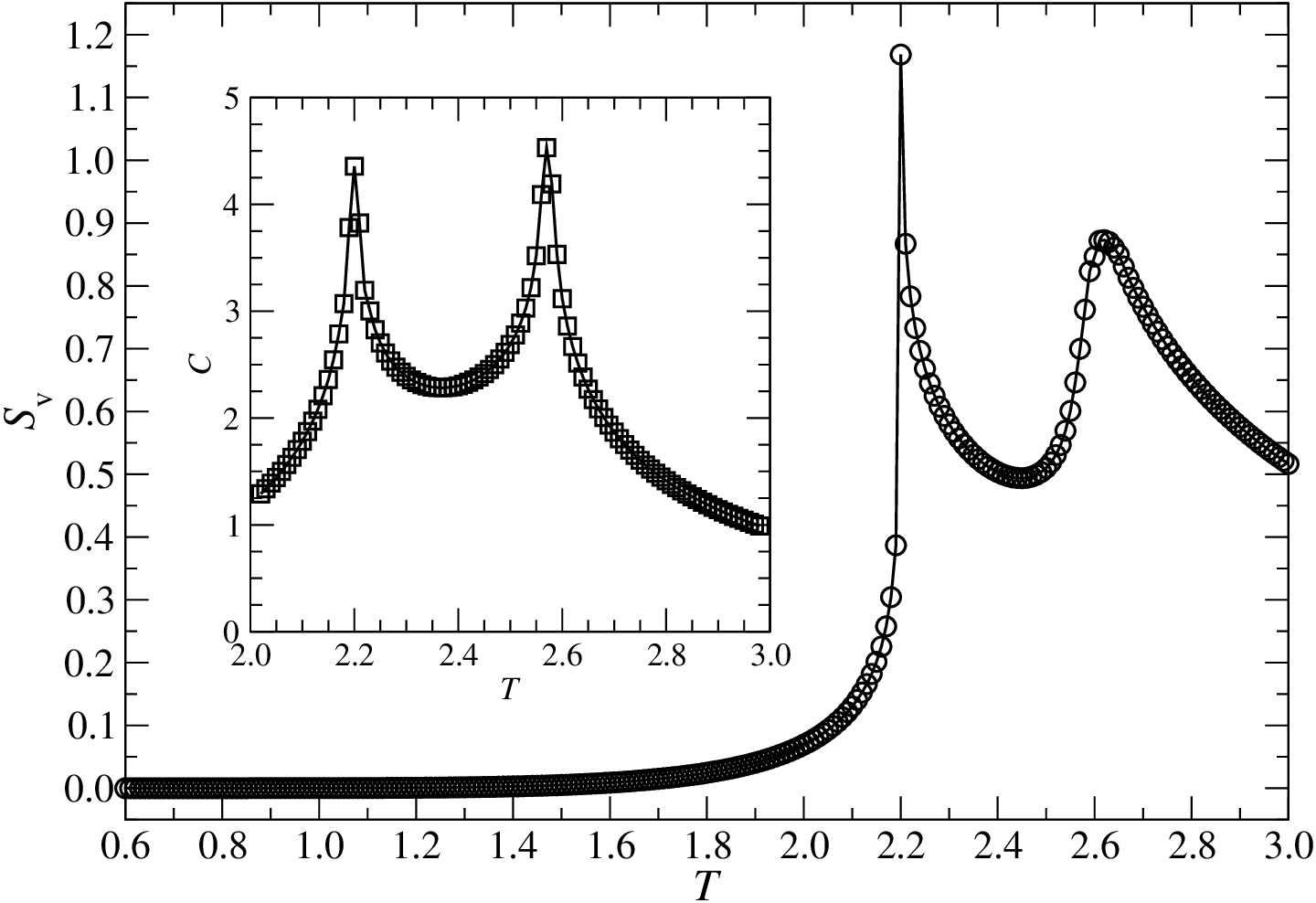}}
\caption{(Color online) The temperature dependence of the entanglement entropy
$S_v$ for $q=2$ and $f=2$. The first maximum in $S_v$ coincides with the critical
temperature $T_{c,1}(2)$ plotted in Fig.~\ref{fig3}, and the second transition
appears at $T_{c,2}(2) = 2.57$. The specific heat, plotted in the inset,
reveals two maxima corresponding to the phase transition temperatures $T_{c,1}(2)$
and $T_{c,2}(2)$.}
\label{fig5}
\end{figure}

The entanglement von Neumann entropy $S_v$ when $q = 2$ is plotted in
Fig.~\ref{fig5}. Evidently, our calculations of $S_v$ result in two maxima,
not only a single maximum as expected for the single phase transition observed
in Fig.~\ref{fig3}. Hence, the entanglement entropy can indicate the existence
of another phase transition, which could not be detected by the sub-site order
parameter $\langle O_{\alpha}\rangle$. The phase transition at lower temperature,
$T_{c,1}(q=2)=2.1973$, coincides with the one plotted in Fig.~\ref{fig3},
whereas the higher-temperature phase transition appears at $T_{c,2}(q=2)=2.57$.
To support this result obtained by $S_v$, we also calculated the specific heat
$C$, as shown in the inset. There are two evident maxima in $C$, which remain
present in our model at the identical critical temperatures $T_{c,1}(2)$ and
$T_{c,2}(2)$. The sub-site order parameter $\langle O_{\alpha}\rangle$ in
Fig.~\ref{fig3} has not reflected the higher-temperature phase transition
at all. Thus, we have achieved a new phase transition point, which is likely
pointing to a topological ordering. A second-order transition has been
found in the out-of-equilibrium Axelrod model~\cite{AxRev}. In addition,
the existence of modulated order parameter with two different phase transition
temperatures has been reported earlier, often being associated with experimental
measurements of the magnetization in crystal alloys~\cite{Ito,Sakon}.

\begin{figure}[tb]
\centerline{\includegraphics[width=0.75\textwidth,clip]{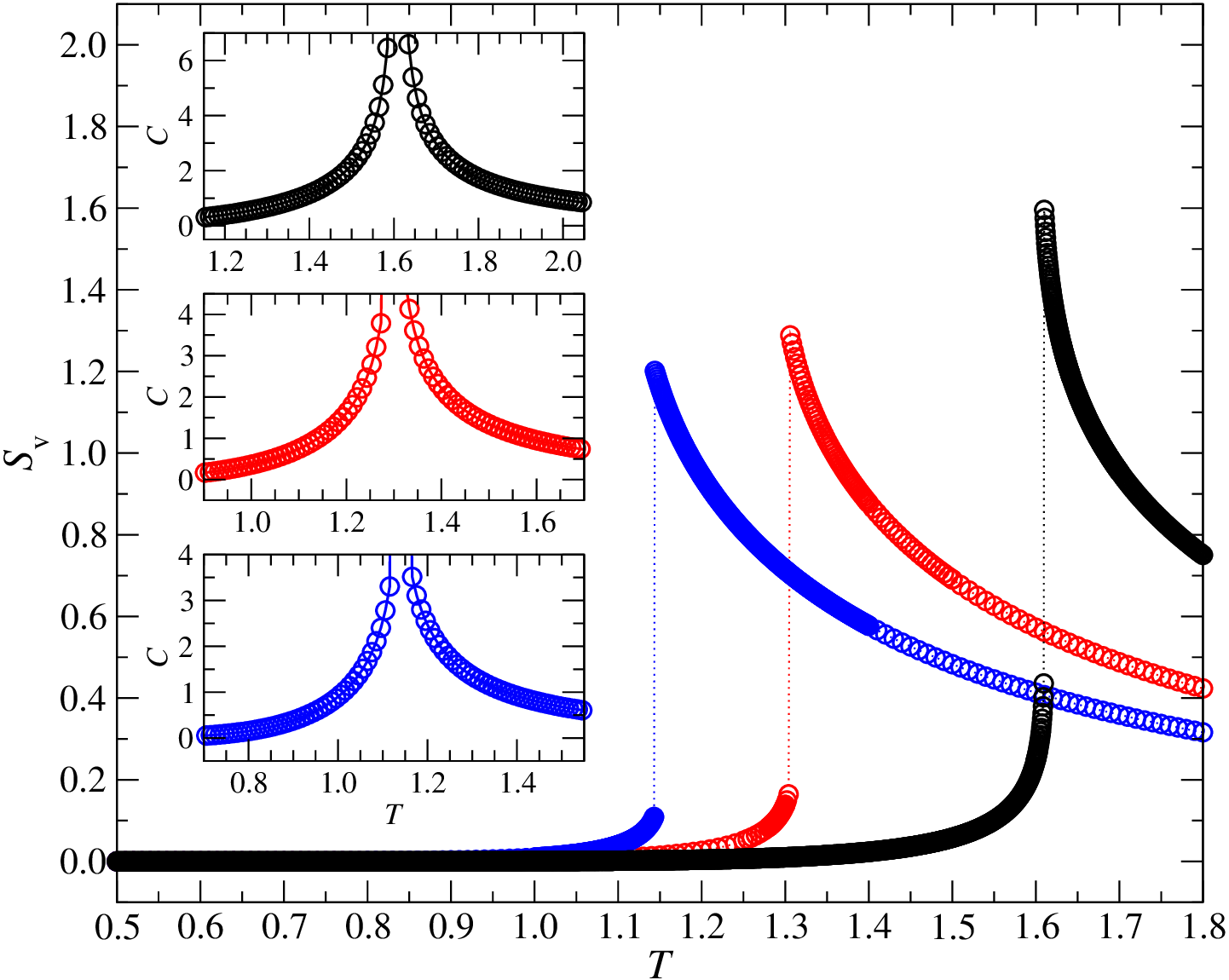}}
\caption{(Color online) The entanglement von Neumann entropy for $q=3$ (black),
$q=4$ (red), and $q=5$ (blue) shows a single maximum when $f=2$. Inset: the
specific heat $C$ also reflects the single (first-order) phase transition
temperature.}
\label{fig6}
\end{figure}

The entanglement entropy $S_v$ exhibits a single maximum for any $q>2$ as seen in
Fig.~\ref{fig6}. The discontinuity of $S_v$ at the phase transition temperature
$T_t(q)$ is characteristic for the first order phase transition. The three
insets display the specific heat with the single maximum for each $q>2$ at the
transition temperature, which is in full agreement with the sub-site order parameter.
Therefore, we conclude the existence of the single phase transition point of the
first order if $q>2$.

\begin{figure}[tb]
\centerline{\includegraphics[width=0.75\textwidth,clip]{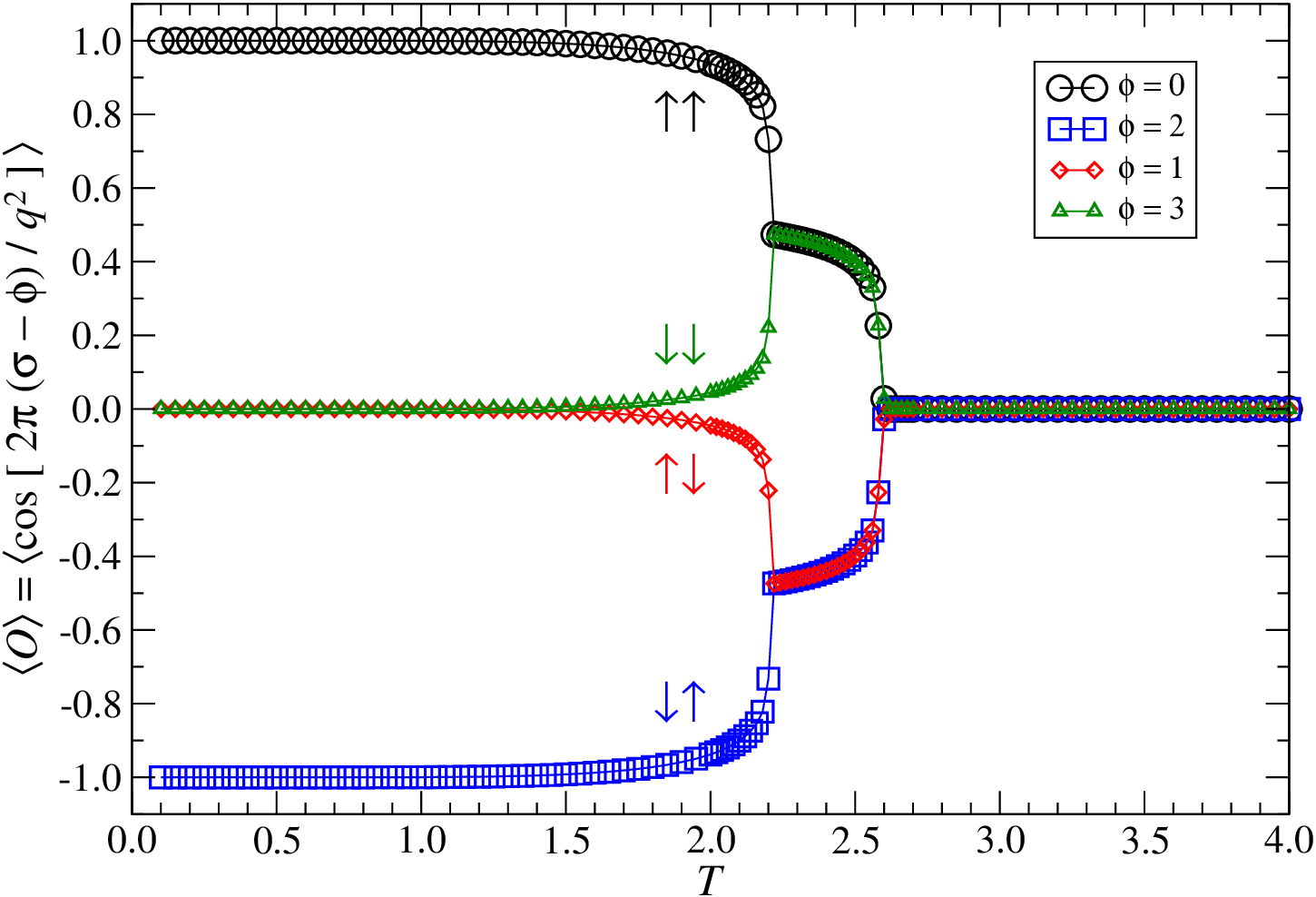}}
\caption{(Color online) The complete order parameter acting on the $q^2$-state
variable $\xi$ exhibits the presence of the two phase transition temperatures
if $q=2$ and $f=2$. All of the four reference spin levels (labeled by $\phi$)
are displayed after the spontaneous symmetry breaking occurs.}
\label{fig7}
\end{figure}

Figure~\ref{fig7} shows the complete order parameter when $q=2$ as defined in
Eq.~\eqref{order2}. Obviously, the non-analytic behavior of $\langle O\rangle$
points to the two distinguishable critical temperatures $T_{c,1}(2)$ and
$T_{c,2}(2)$, which completely coincide with the critical temperatures depicted
in Fig.~\ref{fig5}. Since the $q^2$-state spin $\xi$ has four degrees of freedom,
by targeting the parameters $\phi=0,1,2,3$ separately, the complete order
parameter is explicitly evaluated. It satisfies the condition that the sum of
all four complete order parameters at any temperature has to be zero. The
mechanism of the spontaneous symmetry breaking at low temperatures causes that
the free energy is four-fold degenerate at most. This is related to the four
equivalent free energy minima with respect to the complete order parameter.
Accessing any of the four free energy minima is numerically feasible by
targeting the reference spin state $\phi$.

Let us denote the four spin state at the vertex by the notation
$|\sigma^{(1)}\sigma^{(2)}\rangle$.
There are four possible scenarios for the order parameter $\langle O\rangle$
as shown in Fig.~\ref{fig7}. These scenarios are depicted by the black circles
($\phi=0$), the red diamonds ($\phi=1$), the blue squares ($\phi=2$), and the
green triangles ($\phi=3$), which correspond to the following vertex
configurations $|\hspace{-0.12cm}\uparrow\uparrow\rangle$,
$|\hspace{-0.12cm}\uparrow\downarrow\rangle$,
$|\hspace{-0.12cm}\downarrow\uparrow\rangle$,
and $|\hspace{-0.12cm}\downarrow\downarrow\rangle$, respectively.

At zero temperature there are three minima of the free energy leading to
the three different complete order parameters $\langle O\rangle$ being $-1$,
$0$, and $+1$. There are four minima of the free energy if $0<T<T_{c,1}(2)$
so that the order parameter has four different values $\langle O\rangle =
-1+\varepsilon$, $-\varepsilon$, $+\varepsilon$, and $+1-\varepsilon$ with
the condition $0<\varepsilon\leq\frac{1}{2}$. It means the two states share
the same free energy minimum when the order parameter is zero at $T=0$ and
$\varepsilon=0$. In the temperature interval $T_{c,1}(2)\leq T< T_{c,2}(2)$,
there are only two free energy minima present and the order parameter pair
for $\phi=0$ and $\phi=3$ becomes identical as well as the pair for $\phi=1$
and $\phi=2$. The only single free energy minimum is resulted at $T\geq
T_{c,2}(2)$ when the order parameter is zero, which is typical for the
disordered phase.

Let us stress that at the temperatures in between
$T_{c,1}(2)$ and $T_{c,2}(2)$, the pair of the site configurations
$|\hspace{-0.12cm}\uparrow\uparrow\rangle$ and
$|\hspace{-0.12cm}\downarrow\downarrow\rangle$
is indistinguishable by the complete order parameter
(i.e. the black and green symbols coincide), and the same topological
uniformity happens for the pair of the site configurations
$|\hspace{-0.12cm}\uparrow\downarrow\rangle$ and
$|\hspace{-0.12cm}\downarrow\uparrow\rangle$. In other words,
the anti-parallel alignments between the spins $\sigma^{(1)}$ and
$\sigma^{(2)}$ are preferable in the temperature region
$T_{c,1}(2)\leq T< T_{c,2}(2)$.

Notice that if the critical exponent $\beta$ of the complete order parameter
is calculated at the critical temperatures $T_{c,1}(2)$ and $T_{c,2}(2)$, we
found out that $\beta\approx\frac{1}{18}$ if $T\to T_{c,1}(2)$, whereas the
other exponent remains identical as discussed earlier, in particular,
$\beta\approx\frac{1}{9}$ if $T\to T_{c,2}(2)$.

\begin{figure}[tb]
\centerline{\includegraphics[width=0.75\textwidth,clip]{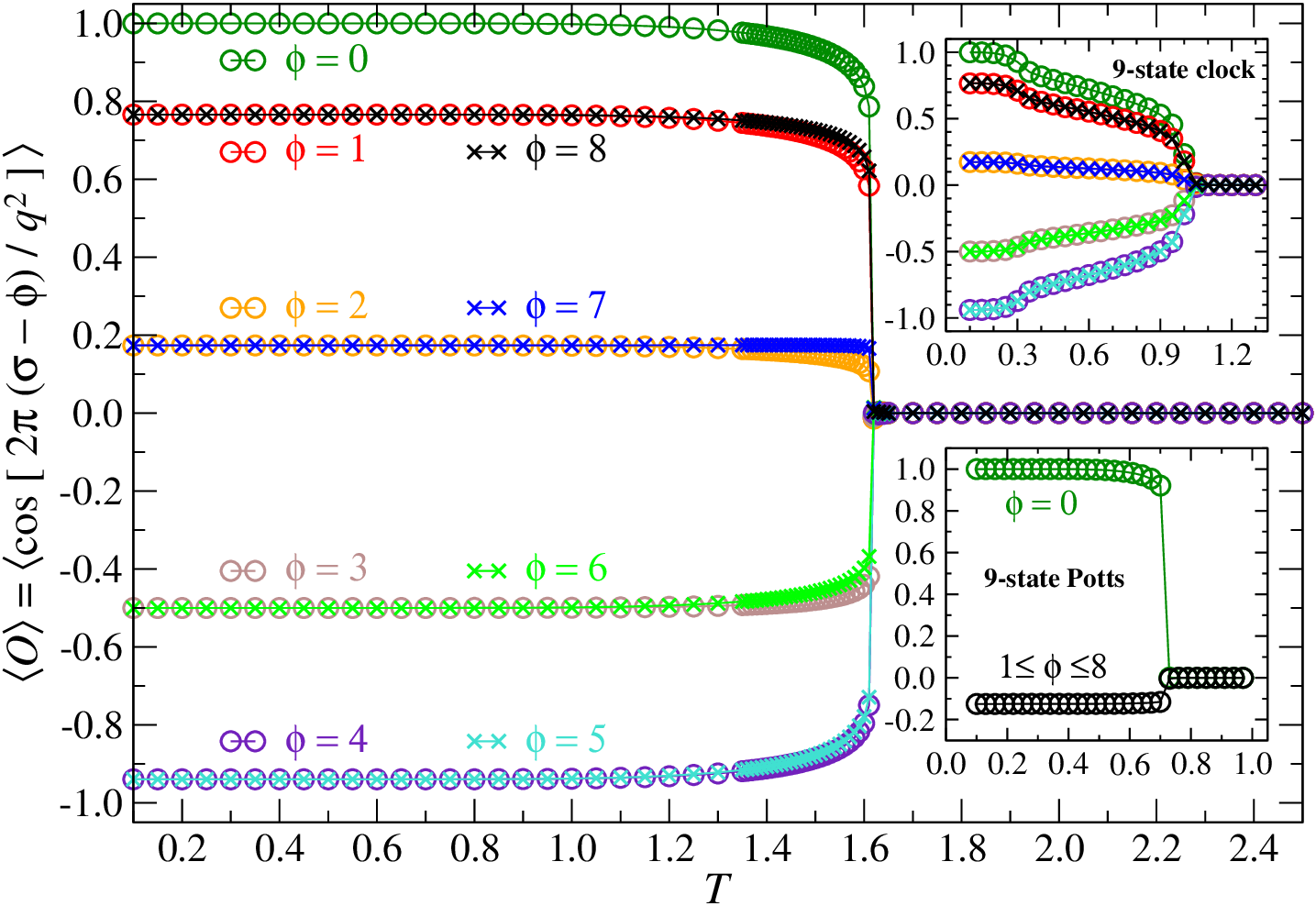}}
\caption{(Color online) The complete order parameter acting on the $q^2$-state
spin $\xi$ for $q=3$ if plotted with respect to the nine reference levels
$\phi=0,1,...,8$ ($f=2$). For comparison, the two insets show the order
parameter of the standard $9$-state clock model (the upper panel) and the
standard $9$-state Potts model (the lower panel) indexed by the reference
levels $\phi$.}
\label{fig8}
\end{figure}

In the same analogy, we plotted the complete order parameter for $q=3$ in
Fig.~\ref{fig8}. The free energy is five-fold degenerated at zero temperature
unless the symmetry breaking mechanism (enhanced by $\phi$) selects one of them.
This mechanism results in the five distinguishable order parameters within
$0\leq\phi\leq8$, which decouple into nine different order parameters when
$0<T<T_t(3)$. Just a single free energy minimum is characteristic in the
disordered phase at $T\geq T_t(3)$ exhibiting a uniform $\langle O\rangle=0$.
In order to compare the main differences of the complete order parameter
between our model and the standard $9$-state clock model or the $9$-state
Potts models, we plotted the respective order parameter in the insets of
Fig.~\ref{fig8}. In the former case (the clock model) there are always
five distinguishable order parameters originating in the five-fold
degeneracy of the free energy, and the order parameters in our model and the
$9$-state clock model are identical at $T=0$ only. However, the five-fold
degeneracy remains within the interval $0<T<T_t(3)$. (We remark that the BKT
phase transitions~\cite{BKT1,BKT2} of the infinite order is present in the
$q\geq5$-state clock models~\cite{BKT3}.) In the latter case (the Potts model),
there are only two distinguishable order parameters out of nine below the phase
transition point. (Recall that the total sum of $\langle O\rangle$ over all
$\phi$ is always zero. The discontinuity in the complete order parameter at
$T_t(3)$ in our model and the $9$-state Potts model reflects the first-order
phase transition~\cite{FYWu}.

\begin{figure}[tb]
\centerline{\includegraphics[width=0.75\textwidth,clip]{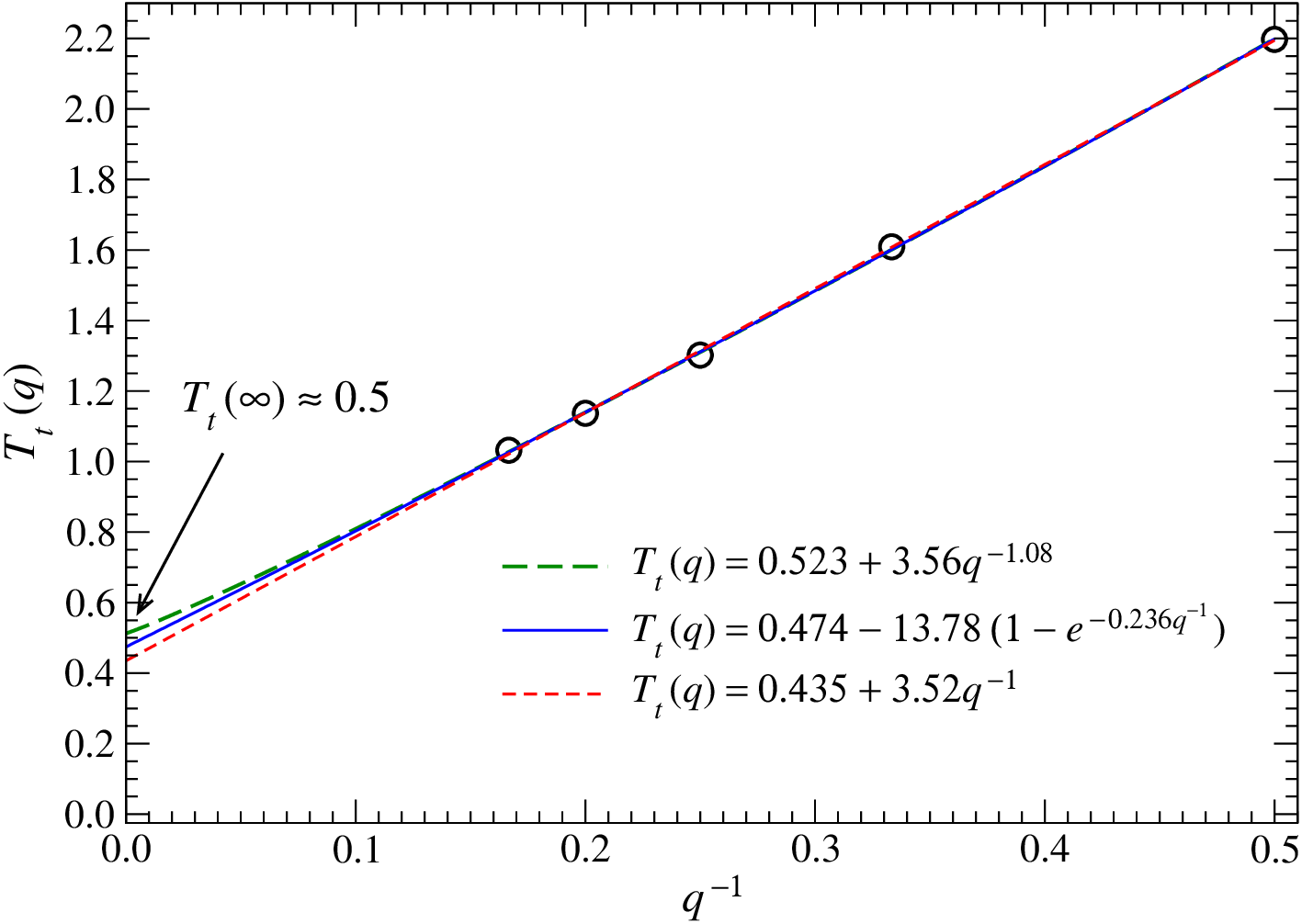}}
\caption{(Color online) The there variants of the extrapolated transition
temperature $T_t(q\to\infty)$ by the power-law fitting (the green long-dashed
line), the exponential fitting (the blue full line), and the inverse
proportionality (the red short-dashed line).}
\label{fig9}
\end{figure}

If the number of the spin degrees of freedom $q$ is extrapolated toward
the asymptotic limit, $q\to\infty$, a non-zero phase transition temperature
$T_t(\infty)$ is resulted. We carried out the three independent extrapolations
as depicted in Fig.~\ref{fig9} by means of the least square fitting. In
particular, the power-law $T_t(q)=T_t(\infty)+a_0\,q^{-a_1}$, the exponential
$T_t(q)=T_t(\infty)+a_0 (1-e^{-a_1/q})$, and the inverse proportional
$T_t(q)=T_t(\infty)+a_0\, q^{-1}$ fitting functions were used to obtain
$T_t(\infty)$, $a_0$, and $a_1$ parameters. All of them yielded the
non-zero transition temperature $T_t(\infty)\approx0.5$. Out of these
findings, we conjecture the existence of the ordered phase, i.e.,
the non-zero phase transition temperature $T_t(q)$ persists for any
$q\geq2$.

\section{Discussion and conclusion}

Having been motivated by the Axelrod model, we studied a multi-state
thermodynamic spin model we proposed for this purpose. Our model is
defined on the two-dimensional infinite square lattice. The spin model
is analyzed by the numerical tools of the statistical physics in order
to calculate equilibrial properties over a social model. We focused on
analyzing phases and the phase transitions. A similar thermodynamic
Axelrod model with $q$-state Potts interactions has been first studied
analytically by Gandica {\it et al.}~\cite{Gandica} on one-dimensional
chains only. In this case the typical thermodynamic properties of
one-dimensional interacting spin systems were concluded with no phase
transition, i.e., the phase transition occurs at $T=0$ only. We have
reproduced features of the $q^2$-state Potts model and we have carried
out numerical analyses on the identical Hamiltonian of the thermodynamic
Axelrod model, as defined in Ref.~\cite{Gandica}, for $q=2$ and $q=3$,
with the respective non-zero phase transition temperature on the
two-dimensional square lattice {\it cf.} Fig.~\ref{fig2}.

The original out-of-equilibrium Axelrod model in one dimension in the
presence of the noise behaves like an ordinary thermodynamic
one-dimensional interacting particle system described by the Potts-like
Hamiltonian~\cite{Gandica}. As demonstrated in Ref.~\cite{Toral1}, the
order-disorder transition induced by the noise depends on the value of
$q$, with the dependence becoming weaker for larger values of $q$. In the
cited paper, the system tends to homogeneity if the noise is low (because of
unstable disordered configurations with respect to the noise perturbations)
 and vice versa, the system prefers heterogeneity whenever the noise rate
gets strong (disappearance of domains is compensated by creating the
new ones). We considered the case of $f=2$ in our model, where we have
observed a strong $q$-dependence. This affects the phase transition
temperature (the critical noise), which decreases with increasing $q$.
Moreover, we have provided the thermodynamic analog of the Axelrod model
in two dimensions, cf. Fig.~2. In addition, our model does not contain
the Potts interactions only, but includes the clock model interactions
with a richer communication structure (interaction).
Such multi-spin model is again mapped onto mutually communicating
individuals subject to a noise, which prevents them in communication. The
gradual increase of the noise disables the formation of larger clusters of
the individuals who share specific cultural features, e.g., interests
(the cluster size is quantified by the order parameter). The raising
noise suppresses correlations at longer distances and exhibits the same
character as the thermal fluctuations. Each individual is characterized
by two independent features ($\alpha$), and each feature assumes
$q$ different traits (interests) resulting in $q^2$ cultural settings
of each individual.

We have found out that such social system exhibits two phase transitions
when $q=2$. Using the above-mentioned examples, one can interpret results
in the following: let, for instance, the first feature describe the two
activities: `reading of books' ($\sigma^{(1)}=\uparrow$) and `listening to
music' ($\sigma^{(1)}=\downarrow$), whereas the second two-state feature
involves `manual activity' ($\sigma^{(2)}=\uparrow$) and `intellectual
activity' ($\sigma^{(2)}=\downarrow$). Both of the phase transitions are
continuous separating three phases, which are classified into the (i)
low-noise regime, (ii) the medium-noise regime, and (iii) the high-noise
regime.

(i) In the low-noise regime, the individuals tend to form a single dominant
cluster, where the complete order parameter has four values (three if $T=0$
only), see Fig.~\ref{fig7}.  The statistical probability of forming the
dominant clusters is proportional to $\langle O\rangle$. If the noise
increases and the complete order parameter decreases to the values of
$\langle O\rangle=\frac{1}{2}$, the four different clusters are formed,
and a final size of the dominant cluster (chosen be setting the parameter
$\phi$) decreases proportionally to this complete order parameter. (ii)
In the medium-noise regime, an interesting topological regime reveals just
two equally likely traits of the individuals. In the social terms, the
pairing of the cultural settings coincides either with (1) the equal mixture
of those individuals who `read books' and `do manual activity'
($\uparrow\uparrow$) and the individuals who `listen to music' and `do
intellectual activity' ($\downarrow\downarrow$) or (2) the equal mixture
of those who `listen to music' and `do manual activity' ($\downarrow\uparrow$)
and those who `read books' and `do intellectual activity' ($\uparrow\downarrow$).
(iii) In the high-noise regime, the clusters are not significant (the correlation
length decreases to zero if the noise increases), and the individuals behave in
a completely uncorrelated way.

A discontinuous phase transition of the first order is present when
the number of the traits $q>2$. In the low-noise regime, larger clusters
of individuals with a given cultural setting (out of $q^2$) are formed.
The selected cultural setting of the dominant cluster sizes is proportional
to the order parameter $\langle O\rangle$. This is equivalent to the ordered
multi-state spin phase below the phase transition noise $T_t(q)$. The regime
of the uncorrelated individuals (disordered phase) appears above the phase
transition noise. The low-noise regime is separated from the high-noise
regime by a discontinuity of the cluster size (the complete order parameter).

In the asymptotic limit of the number of the traits (the cultural settings),
the extrapolation of the phase transition noise results in the non-zero
$T_t(\infty)$. We conjecture that the phase transition noise $T_t(\infty)$
remains finite (being approximately $0.5$). We interpret this result as the
permanent existence of the correlated clusters below the non-zero phase
transition point $T_t(\infty)$.

\section{Acknowledgments}
This work was supported by the grants QIMABOS APVV-0808-12, VEGA-2/0074/12, and
EU project SIQS No. 600645.

\end{document}